\documentclass[aps,prb,showpacs,twocolumn]{revtex4-1}
\usepackage{amssymb}
\usepackage{amsmath}
\usepackage{graphicx}
\usepackage{epsfig}
\usepackage{subfigure}
\usepackage{appendix}

\begin{document}

\title{Manifestation of a topological gapless phase in a two-dimensional chiral symmetric system through Loschmidt echo}
\author{K. L. Zhang}
\author{Z. Song}
\email{songtc@nankai.edu.cn}
\affiliation{School of Physics, Nankai University, Tianjin 300071, China}

\begin{abstract}
Unlike the edge state of a topological insulator where its energy level lives in the bulk energy gap, the edge state of a topological semimetal hides in the bulk spectrum and is difficult to be identified by the energy. We investigate the sensitivity of bulk and edge states of the gapless phase for a topological semimetal to the disordered perturbation via a concrete two-dimensional chiral symmetric lattice model. The topological gapless
phase is characterized by two opposite vortices in the momentum space and nonzero winding numbers, which correspond to the edge flatband when the open boundary condition is applied. For this system, numerical results reveal that a distinguishing feature is that the robustness of the edge states against weak disorder and the flatband edge modes remain locked at zero energy in the presence of weak chiral-symmetry-preserving disorder. We
employ the Loschmidt echo (LE) for both bulk and edge states to study the dynamic effect of disordered perturbation. We show that, for an initial bulk state, the LE decays exponentially, whereas it converges to a constant for an initial edge state in the presence of weak disorder. Furthermore, the convergent LE can be utilized to identify the positions of vortices as well as the phase diagram. We discuss the realization of such dynamic
investigations in a topological photonic system.
\end{abstract}

\maketitle

\section{Introduction}

Topological state of matter \cite{Hasan,XLQ,CKC,HMW} have become the focus
of intense research in many branches of physics and provides a fertile
ground for demonstrating the concepts in high-energy physics, including
Majorana \cite{LF,RML,VM,SNP,YO,NR}, Dirac \cite{AHCN,ZKL,ZKL2,JAS,ZW,JX,SMY}
and Weyl fermions \cite{MH,SMH,BQL,BQL2,CS,XW,HW,SYX,SYX2}.\ These concepts
relate to topological gapless phases and corresponding edge modes, not only
exhibiting new physical phenomena with potential technological applications,
but also deepening our understanding on state of matters. System in the
topological gapless phase exhibits band structures with band-touching points
in the momentum space, where these kinds of nodal points appear as
topological defects of Bloch vector field. On the other hand, a gapped phase
can be topologically non-trivial, commonly referred to as topological
insulators and superconductors. Such phase is associated at least with two isolated bulk energy bands, where the band structure of each is
characterized by nontrivial topological index. A particularly important
concept is the bulk-boundary correspondence, which links the nontrivial
topological invariant in the bulk to the localized edge modes. In general,
edge states are the eigenstates of Hamiltonian that are exponentially
localized at the boundary of the system. A gapped topological phase is
always associated with an bulk energy gap, while a topological
gapless phase, commonly referred to as topological semimetals
and nodal superconductors, can exhibit topological protected Fermi points or
nodal points (We refer to the bulk energy gap as the energy gap of the system with translational symmetry and without disorder throughout this paper). Accordingly, unlike the edge state of a topological insulator, where its energy level lives in the bulk energy gap,
the edge state of a topological gapless phase hides in the bulk spectrum,
and is hard to be identified by the energy. These edge states can form
partial flat band in a ribbon geometry \cite{FM,RS,YW}, which also exhibit
robustness against disorder \cite{WK}. Recently, it has been pointed that
Majorana zero modes are not only attributed to topological superconductors.
A two-dimensional (2D) topologically trivial superconductors without chiral
edge modes can host robust Majorana zero modes in topological defects \cite%
{ZBY,YZB,WQY}. In experimental aspect, photonic systems provide a convenient
and versatile platform to design various topological lattice models and
study different topological states \cite{LuL,OzawaT}.

In this paper, we investigate the sensitivity of bulk and edge states of
gapless phase for topological semimetal to the disordered
perturbation via a concrete two-dimensional chiral symmetric lattice model.
We employ the Loschmidt echo (LE) for both bulk and edge states to study the
dynamic effect of disordered perturbation. The LE is a measure of the
revival occurring when an imperfect time-reversal procedure is applied to a
complex quantum system. It allows to quantify the sensitivity of quantum
evolution to perturbations. This work aims to shed light on the nature of
topological edge modes associated with gapless phase for topological semimetal with chiral symmetry, rather than gapped topological
materials. We show that for a initial bulk state LE decays exponentially,
while converges to a constant for an initial edge state in the presence of
weak chiral-symmetry-preserving disorder. Our results provide a
dynamic way to identify topological edge states arising from topological
gapless phase in 2D chiral symmetric system. The reason is that unlike the
edge states in topological insulator, here the edge flat band hides in a
continuous spectrum. There is no bulk energy gap to protect the
channel of the edge states. Thanks to the photonic system, where the Pauli
exclusion not obeyed, a single-particle state can be amplified by the large
population of photons. The phase diagram can be detected by using edge-state
photon dynamics.

This paper is organized as follows. In Sec. \ref{Anderson localization and
Loschmidt echo}, we present the introduction of Loschmidt echo and the idea
about apply it to the bulk and edge states in gapless systems. In Sec. \ref%
{Model without disorder}, we introduce a square lattice without disorder to
illustrate our method. Section \ref{Dynamic detection of edge modes} focus
on the dynamics of the system in the presence of disorder, and demonstrates
the dynamics method of detect edge modes. Finally, our conclusion and
discussion are given in Sec. \ref{Discussion}.

\section{Edge states and Loschmidt echo}

\label{Anderson localization and Loschmidt echo} Anderson localization is a
basic condensed matter physics phenomenon, which describes the absence of
diffusion of waves in a disordered medium \cite{PWAnderson}. It turns out
that particle localization is possible in a lattice potential, provided that
the strength of disorder in the lattice is sufficiently large. The
confinement of waves in a disordered medium has been observed for
electromagnetic \cite{AAChabanov,TSchwartz} and acoustic \cite{HHu} waves in
disordered dielectric structures, and for electron waves in condensed
matter. On the other hand, to capture the effect of disorder on the
dynamics, one can employ a concept of LE or fidelity. LE is a measure of
reversibility and sensitivity to perturbations of quantum evolutions. An
initial quantum state $|\psi (0)\rangle $ evolves during a time $T$ under a
Hamiltonian $H_{0}$ reaching the state $|\psi (t)\rangle $. Aiming to
recover the initial state $|\psi (0)\rangle $ a new Hamiltonian $H$ is
applied between $T$ and $2T$. Quantity $\left\vert \langle \psi
(0)|e^{-iHt}e^{-iH_{0}t}|\psi (0)\rangle \right\vert ^{2}$\ is induced to
measure the fidelity of this recovery. Perfect recover of $|\psi (0)\rangle $
would be achieved by choosing $H=-H_{0}$. In the context of the present
work, the LE is defined as 
\begin{equation}
M(t)=\left\vert \langle \psi (0)|e^{iH_{\mathrm{D}}t}e^{-iH_{0}t}|\psi
(0)\rangle \right\vert ^{2},  \label{LE_def}
\end{equation}%
where $|\psi (0)\rangle $ is the state of the system at time $t=0$, $H_{0}$
is the Hamiltonian of uniform system, $H_{\mathrm{D}}$ is the Hamiltonian $%
H_{0}$\ under disordered perturbation. For certain topological
non-trivial systems, edge states are robust under a weak symmetry-preserving
disorder, still being localized state \cite{OzawaT, APS2011R, APS2012}. Particularly, we numerically observe that the corresponding eigen energy is locked at zero, as it is shown through a concrete model in Sec. \ref{Dynamic detection of edge modes}. Therefore, considering such a topological system, it is expected that (i) when an initial quantum state $|\psi(0)\rangle $ is a local state at bulk, $M(t)$ could decay
to zero due to the fact that $e^{-iH_{0}t}|\psi (0)\rangle $ and $%
e^{-iH_{\mathrm{D}}t}|\psi (0)\rangle $ diffuse in 2D space in
different ways, (ii) when $|\psi (0)\rangle $ is an edge state of %
$H_{0}$, $M(t)$ could be the constant $1$.

Here we consider the LE for a local state $|\psi (0)\rangle $ 
at the edges. We note that $|\psi (0)\rangle $ is almost
be written as the superposition of the flat band edge modes of $H_{0}$ 
or $H_{\mathrm{D}}$, respectively, which result in
\begin{equation}
H_{0}|\psi (0)\rangle \approx 0,H_{\mathrm{D}}|\psi (0)\rangle \approx 0,
\label{M0}
\end{equation}%
for weak disordered perturbation $H_{\mathrm{D}}-H_{0}\ll H_{0}$. Then we always have
\begin{eqnarray}
M(t) &=&\left\vert \langle \psi (0)|e^{iH_{\mathrm{D}}t}e^{-iH_{0}t}|\psi
(0)\rangle \right\vert ^{2}  \notag \\
&\approx &\left\vert \langle \psi (0)|\psi (0)\rangle \right\vert ^{2}=1.
\label{M1}
\end{eqnarray}%
In the following, we will demonstrate this analysis through a
concrete example.

\section{Model without disorder}

\label{Model without disorder} We focus on a concrete 2D chiral symmetric
model to demonstrate the main idea. In the previous works \cite{ZKLSR, WP1}
we have demonstrated that a topologically trivial superconductor emerges as
a topological gapless state, which support Majorana flat band edge modes.
The quantum state is characterized by two band-degeneracy points with
opposite chirality. In the present work, we directly consider a
tight-binding model with the same structure as the Majorana lattice. In the
following, (i) we present the Hamiltonian and the phase diagram for the
topological gapless phase; (ii) we investigate the topological edge states
with nonzero winding numbers.

\begin{figure}[b]
\centering
\includegraphics[width=0.4\textwidth]{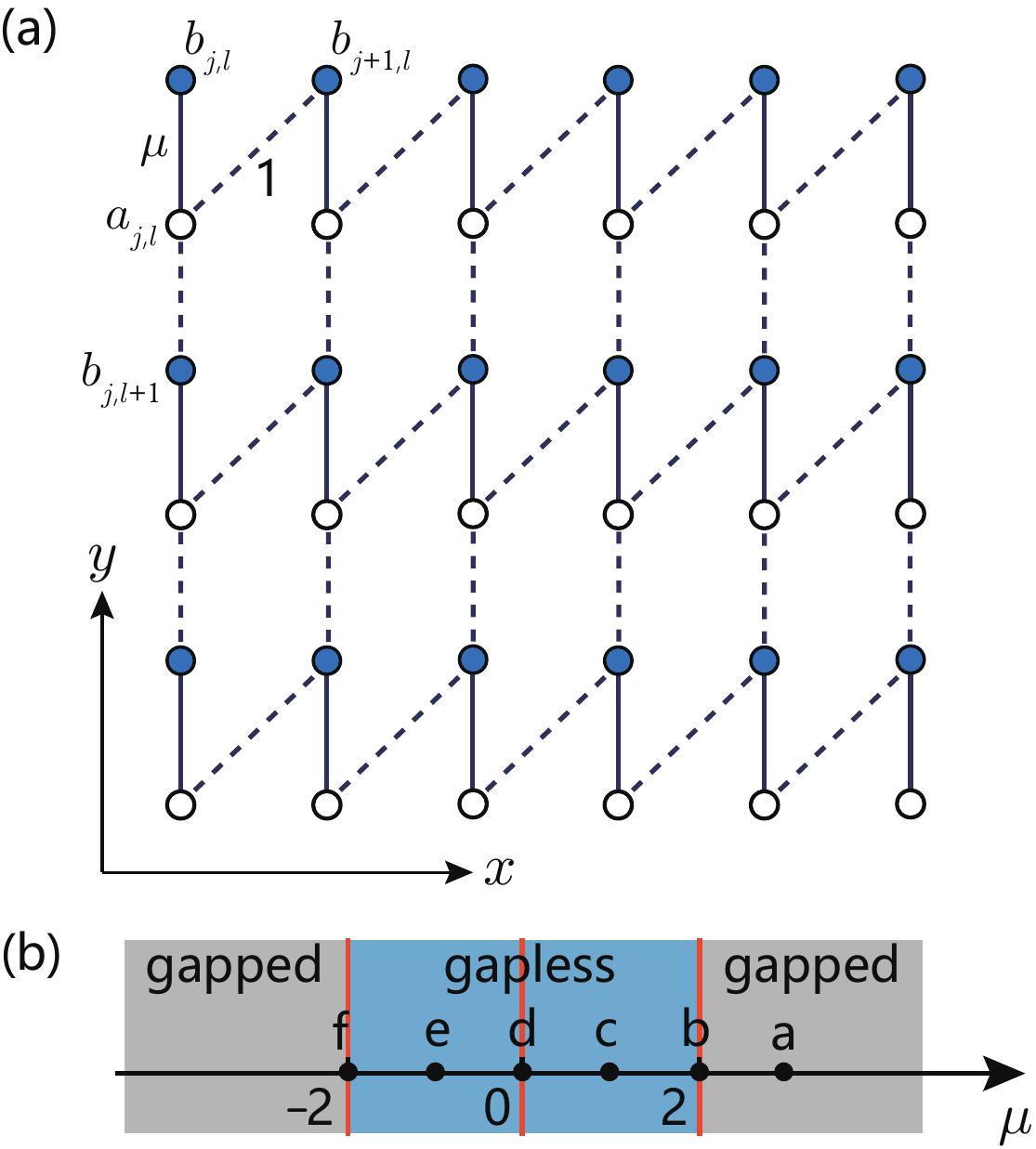}
\caption{(a) Schematic illustration of the 2D tight-binding model with
Hamiltonian Eq. (\protect\ref{H}), which is essentially a bipartite $M\times
N$\ honeycomb lattice. The solid lines and dashed lines represent hopping
terms with strengths $\protect\mu$ and 1, respectively. (b) Phase diagram of
the lattice system with parameter $\protect\mu$. The orange lines indicate
the phase boundary, which separate the topologically trivial gapped phases
(gray) and topological gapless phases (blue). The system with parameter at
the boundary (orange lines) is topologically trivial gapless phase. Points
(a-f) represent the systems in each phases, which are extensively
investigated in Fig. \protect\ref{fig3}.}
\label{fig1}
\end{figure}

\subsection{Model and topological gapless phase}

We consider a tight-binding model on a bipartite $M\times N$ lattice with
the Hamiltonian%
\begin{equation}
H=\sum_{\mathbf{r}}\left( \mu a_{\mathbf{r}}^{\dagger }b_{\mathbf{r}}+a_{%
\mathbf{r}}^{\dagger }b_{\mathbf{r}+\hat{x}}+a_{\mathbf{r}}^{\dagger }b_{%
\mathbf{r}+\hat{y}}\right) +\text{h.c.,}  \label{H}
\end{equation}%
where $\mathbf{r}=\left( j,l\right) $ is the coordinates of lattice sites
and $a_{\mathbf{r}}$ and $b_{\mathbf{r}}$\ are the fermion or boson
annihilation operators at site $\mathbf{r}$ in sublattice of A and B,
respectively. Vectors $\hat{x},$ $\hat{y},$ are the unitary lattice vectors
in the $x$ and $y$ directions. The hopping between neighboring sites is
described by the hopping amplitudes $\mu $ and $1$. The schematic diagram
for the honeycomb lattice is shown in Fig. \ref{fig1}(a). This simple model
can be regard as a strained graphene lattice \cite{VMPereira2009,
HRostami2012, HHPu2013, ASharma2013, DABahamon2013} which is uniaxially
strained along the $y$ direction.

We introduce the Fourier transformations 
\begin{equation}
\left( a_{\mathbf{k}},b_{\mathbf{k}}\right) =\frac{1}{\sqrt{MN}}\sum_{%
\mathbf{r}}\left( a_{\mathbf{r}},b_{\mathbf{r}}\right) e^{-i\mathbf{k}\cdot 
\mathbf{r}}.
\end{equation}%
Then the Hamiltonian with periodic boundary conditions on both directions
can be block diagonalized as%
\begin{equation}
H=\sum_{\mathbf{k}}\left( 
\begin{array}{cc}
a_{\mathbf{k}}^{\dag } & b_{\mathbf{k}}^{\dag }%
\end{array}%
\right) h\left( \mathbf{k}\right) \left( 
\begin{array}{c}
a_{\mathbf{k}} \\ 
b_{\mathbf{k}}%
\end{array}%
\right) ,  \label{2D_bulk}
\end{equation}%
with the core matrix 
\begin{equation}
h\left( \mathbf{k}\right) =\left( 
\begin{array}{cc}
0 & g\left( \mathbf{k}\right) \\ 
g^{\ast }\left( \mathbf{k}\right) & 0%
\end{array}%
\right) ,
\end{equation}%
and $g\left( \mathbf{k}\right) =\mu +\left( e^{ik_{x}}+e^{ik_{y}}\right) .$
We note that the system respects time reversal, chiral, and particle-hole symmetry, i.e., for the Bloch Hamiltonian $h(\mathbf{k})$, we
have $\mathcal{T}h(\mathbf{k})\mathcal{T}^{-1}=h(-\mathbf{k})$, $\mathcal{S}%
h(\mathbf{k})\mathcal{S}^{-1}=-h(\mathbf{k})$, and $\mathcal{C}h(\mathbf{k})%
\mathcal{C}^{-1}=-h(-\mathbf{k})$, with $\mathcal{T=K}$ is the complex
conjugation operator and $\mathcal{S}=\sigma _{z}$, $\mathcal{C}=\sigma _{z}%
\mathcal{K}.$ The core matrix can be written as%
\begin{equation}
h\left( \mathbf{k}\right) =\mathbf{B}\left( \mathbf{k}\right) \cdot \mathbf{%
\sigma },
\end{equation}%
where the components of the Bloch vector $\mathbf{B}\left( \mathbf{k}\right)
=(B_{x},B_{y},B_{z})$\ are%
\begin{equation}
\left\{ 
\begin{array}{l}
B_{x}=\mu +\left( \cos k_{x}+\cos k_{y}\right) \\ 
B_{y}=-\left( \sin k_{x}+\sin k_{y}\right) \\ 
B_{z}=0%
\end{array}%
\right. ,  \label{field}
\end{equation}%
and $\mathbf{\sigma }=\left( \sigma _{x},\sigma _{y},\sigma _{z}\right) $\
are the Pauli matrices. The spectrum is%
\begin{equation}
E_{\mathbf{k}}^{\pm }=\pm \sqrt{\left( \mu +\cos k_{x}+\cos k_{y}\right)
^{2}+\left( \sin k_{x}+\sin k_{y}\right) ^{2}}.  \label{spectrum}
\end{equation}%
We focus on the gapless phase arising from the band degenerate points of the
spectrum. The band degenerate point $\mathbf{k}_{0}=(k_{0x},k_{0y})$ fulfill
the equations%
\begin{equation}
\left\{ 
\begin{array}{l}
\sin k_{0x}+\sin k_{0y}=0 \\ 
\mu +\cos k_{0x}+\cos k_{0y}=0%
\end{array}%
\right. .  \label{para eq}
\end{equation}%
As shown in Fig. \ref{fig3}(a1)-(f1), there have three types of bands
touching configurations: single point, double points, and lines in the $%
k_{x} $-$k_{y}$\ plane, determined by the parameter $\mu $. We are
interested in the non-trivial case (double points) with nonzero $\mu $. Then
from Eq. (\ref{para eq}), we have%
\begin{equation}
k_{0x}=-k_{0y}=\pm \arccos (-\frac{\mu }{2}),  \label{condition eq}
\end{equation}%
in the condition of $\left\vert \mu \right\vert \leqslant 2$. It indicates
that there are two degenerate points for $\mu \neq 0$ and $\left\vert \mu
\right\vert \neq 2$. When $\mu $ vary, the two points move along the line: $%
k_{0x}=-k_{0y}$, and merge at $\mathbf{k}_{0}=(\pm \pi ,\mp \pi )$ or $%
\mathbf{k}_{0}=(0,0)$\ when $\mu =2$ or $\mu =-2$. In the case of $\mu =0$,
the degenerate points become two degenerate lines: $k_{0y}$ $=$ $\pm \pi
+k_{0x}$. The phase diagram is shown in Fig. \ref{fig1}(b) and the bulk
spectra for several typical cases are illustrated in Fig. \ref{fig3}%
(a1)-(f1).

\begin{figure}[tbp]
\centering
\includegraphics[width=0.4\textwidth]{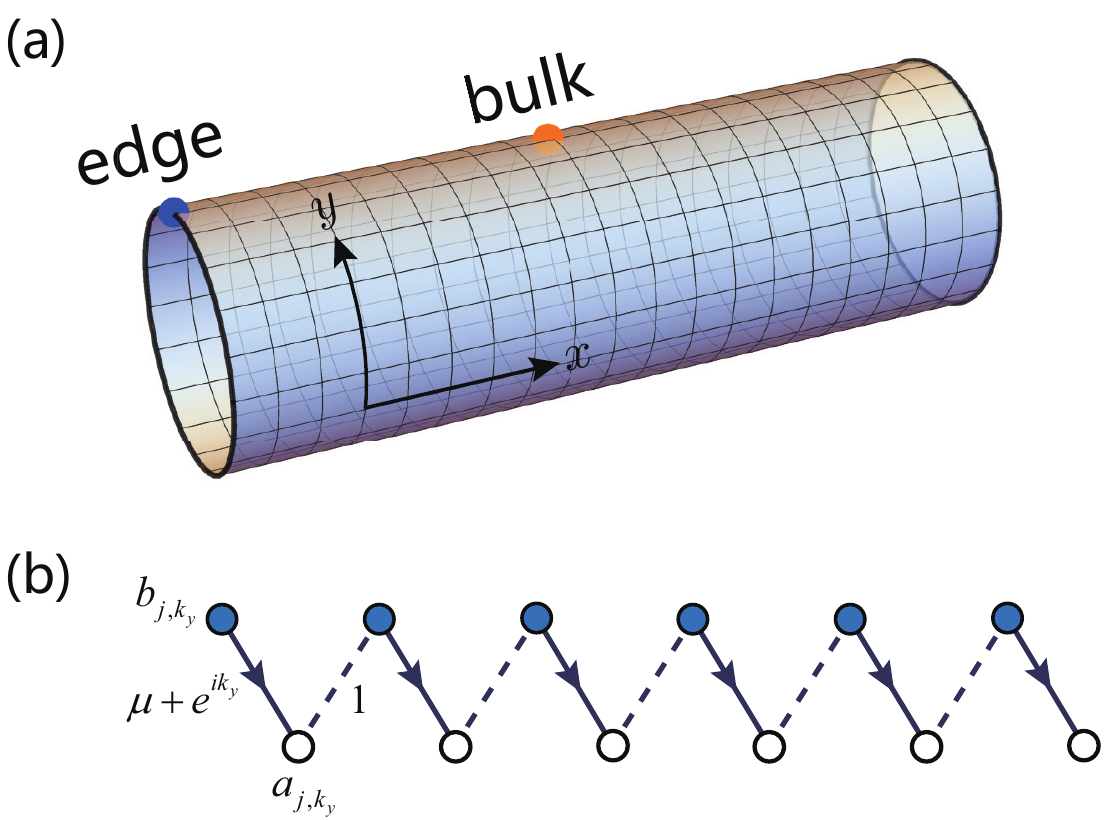}
\caption{(a) Schematic illustration of the geometry of the system with
cylindrical boundary condition. The locations of initial local states at
edge (blue dot) and bulk (orange dot) of the 2D lattice system are
indicated. (b) Schematic illustration of the modified SSH chain $H_{k}$
represented in Eq. (\protect\ref{SSH}). The arrows and dashed lines represent
complex and real hopping terms, $\left( \protect\mu +e^{ik}\right) $ and $1$%
, respectively. The edge modes of a set of modified SSH chains form the flat
band edge modes as bound states located at two edges of the cylinder when
the system is in the blue region of the phase diagram in Fig. \protect\ref%
{fig1}(b).}
\label{fig2}
\end{figure}

\begin{figure*}[tbh]
\centering
\includegraphics[width=1\textwidth]{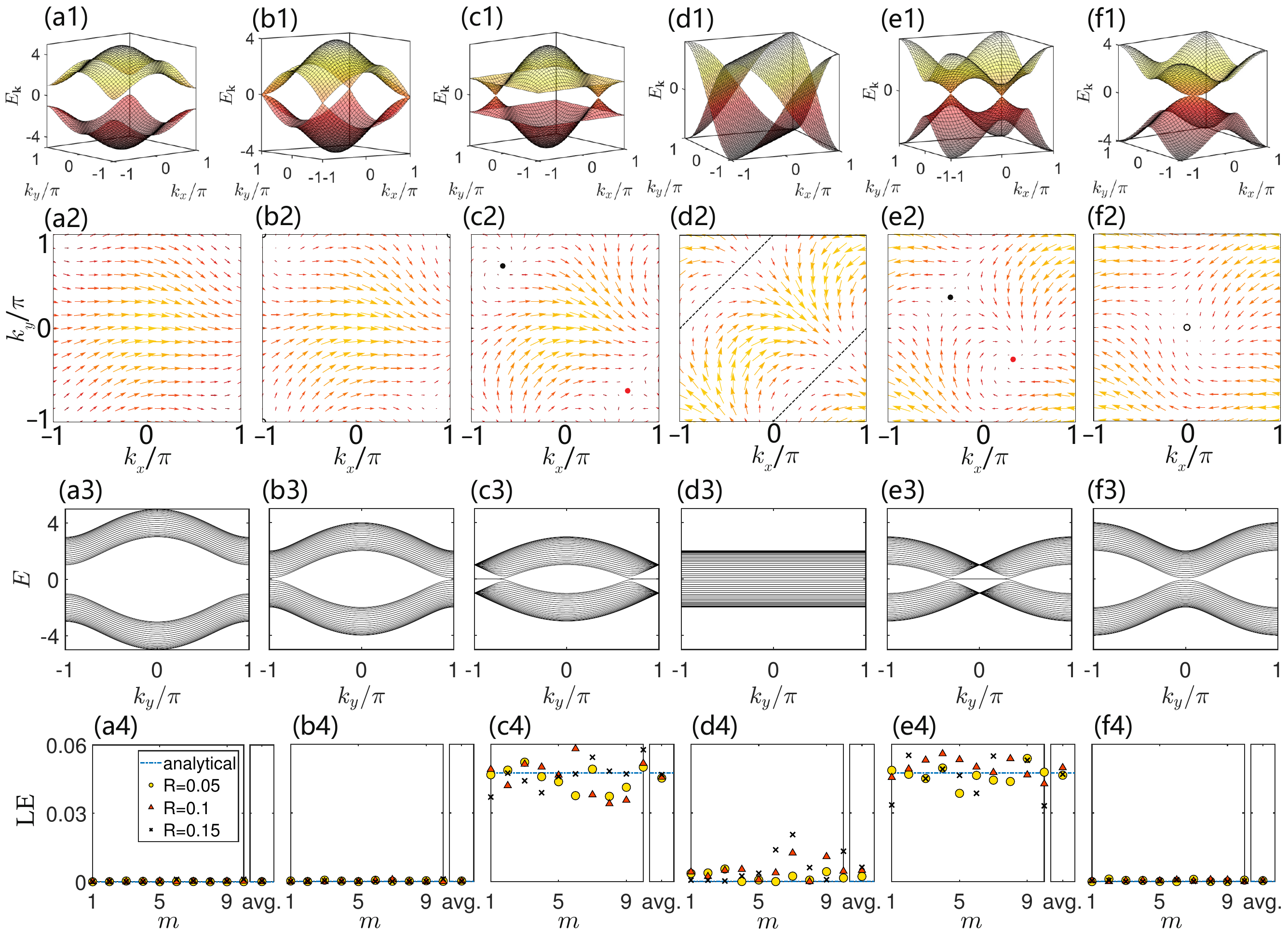}
\caption{Kaleidoscope of quantum phases. (a1)-(f1) Plots of energy spectra
from Eq. (\protect\ref{spectrum}) at six typical points (a-f) marked in the
phase diagram in Fig. \protect\ref{fig1}(b). There, the band structure
exhibits a bulk gap in (a1); a single degeneracy point with
parabolic dispersion in (b1) and (f1); two degeneracy points with linear
dispersion in (c1) and (e1); and two degeneracy lines in (d1). (a2)-(f2)
Plots of Bloch vector field defined in Eq. (\protect\ref{field}) in the
momentum space for six cases corresponding to (a1)-(f1). There are two
vortices in (c2) and (e2) with opposite winding numbers $\pm 1$. As $\protect%
\mu $ increases or decrease, two vortices get close and merge into a single
point in (b2) or (f2), and disappears in (a2). (a3)-(f3) Plots of the
spectra of a set of modified SSH chains (Eq. (\protect\ref{spectrum_H})) in
open boundary condition with $N=40$ for six cases corresponding to
(a1)-(f1). It indicates that the existence of pair of vortices links to a
flat band of the square lattice. (a4)-(f4) Plots of LE obtained by numerical
simulations from Eq. (\protect\ref{LE_def}) at $t=3000J^{-1}$ and analytical
expressions from Eq. (\protect\ref{LE_analytical}), in which the initial
state is taken as the site-state at the edge. Here $J$ is the scale of the
Hamiltonian and we take $J=1$. $R$ is the disorder strength and $m$ denotes
the measurement index. The average value of LEs are plotted in the right of
each panel. The size of the system is $M\times N=80\times80$.}
\label{fig3}
\end{figure*}

The gapless phase of this model can be protected by a $\mathbb{Z}
$-type invariant according to the classification topological semimetals \cite%
{CKC, CKChiu2014}. For isolated band touching point, the topological nature
of the band degeneracy can be considered as a vortex in the momentum space
with integer winding numbers, which is equivalent to concept of the Berry
flux\cite{FHaldane,KSun}. The Berry flux is defined as the contour integral
of the Berry connection in the momentum space \cite{EIBlount,FHaldane}. A
band degenerate point can be regarded as a topological defect and the
topological index can be extracted from the expression of Bloch vector $%
\mathbf{B}\left( \mathbf{k}\right) $. Actually, in the vicinity of the
degenerate points, the Bloch vector can be expressed as the from%
\begin{equation}
\left\{ 
\begin{array}{l}
B_{x}=-\sin k_{0x}\left( q_{x}-q_{y}\right) \\ 
B_{y}=-\cos k_{0x}\left( q_{x}+q_{y}\right) \\ 
B_{z}=0%
\end{array}%
\right. ,  \label{B field}
\end{equation}%
where $\mathbf{q=k-k}_{0}$ is the momentum in another frame and $\mathbf{k}%
_{0}=\left( k_{0x},k_{0y}\right) $ satisfy Eq. (\ref{condition eq}). Around
these degenerate points, the core matrix $h(\mathbf{k})$ can be linearized as%
\begin{equation}
h(\mathbf{q})=\sum_{i,j=1}^{2}c_{ij}q_{i}\sigma _{j},
\end{equation}%
which is equivalent to the Hamiltonian for 2D massless relativistic
fermions. Here $\left( q_{1},q_{2}\right) =\left( q_{x},q_{y}\right) $, $%
\left( \sigma _{1},\sigma _{2}\right) =\left( \sigma _{x},\sigma _{y}\right) 
$ and $c=\left( 
\begin{array}{cc}
-\sin k_{0x} & -\cos k_{0x} \\ 
\sin k_{0x} & -\cos k_{0x}%
\end{array}%
\right) $. The corresponding chirality for these particle is defined as 
\begin{equation}
w=\mathrm{sgn}[\det \left( c\right)]=\mathrm{sgn}[\sin \left(
2k_{0x}\right)],
\end{equation}%
\begin{figure*}[tbh]
\centering
\includegraphics[width=1\textwidth]{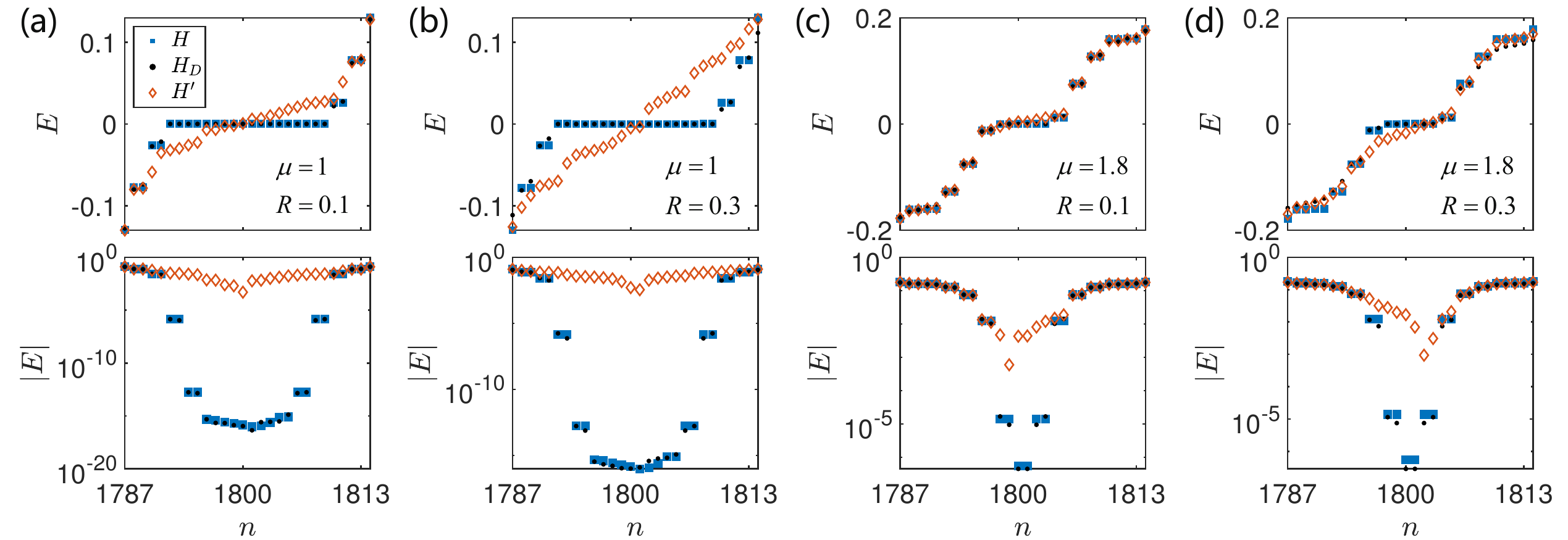}
\caption{Linear and log scale plots of eigenenergy around zero for $H$, $%
H_{D}$ and $H^{\prime}$ with cylindrical boundary condition. $n$ denotes the
sorting index. The parameters are $\protect\mu=1$ for (a) and (b); $\protect%
\mu=1.8$ for (c) and (d); $R=0.1$ for (a) and (c); $R=0.3$ for (b) and (d).
It indicates that the number of zero modes is dependent on $\protect\mu$,
and the zero modes remain unchanged in the presence of
chiral-symmetry-preserving disordered perturbation, while do not survive
under the chiral-symmetry-breaking disordered perturbation. The results are
obtained by numerical diagonalization for the system with $M\times
N=60\times 60$.}
\label{fig4}
\end{figure*}
\begin{figure}[tbh]
\centering
\includegraphics[width=0.45\textwidth]{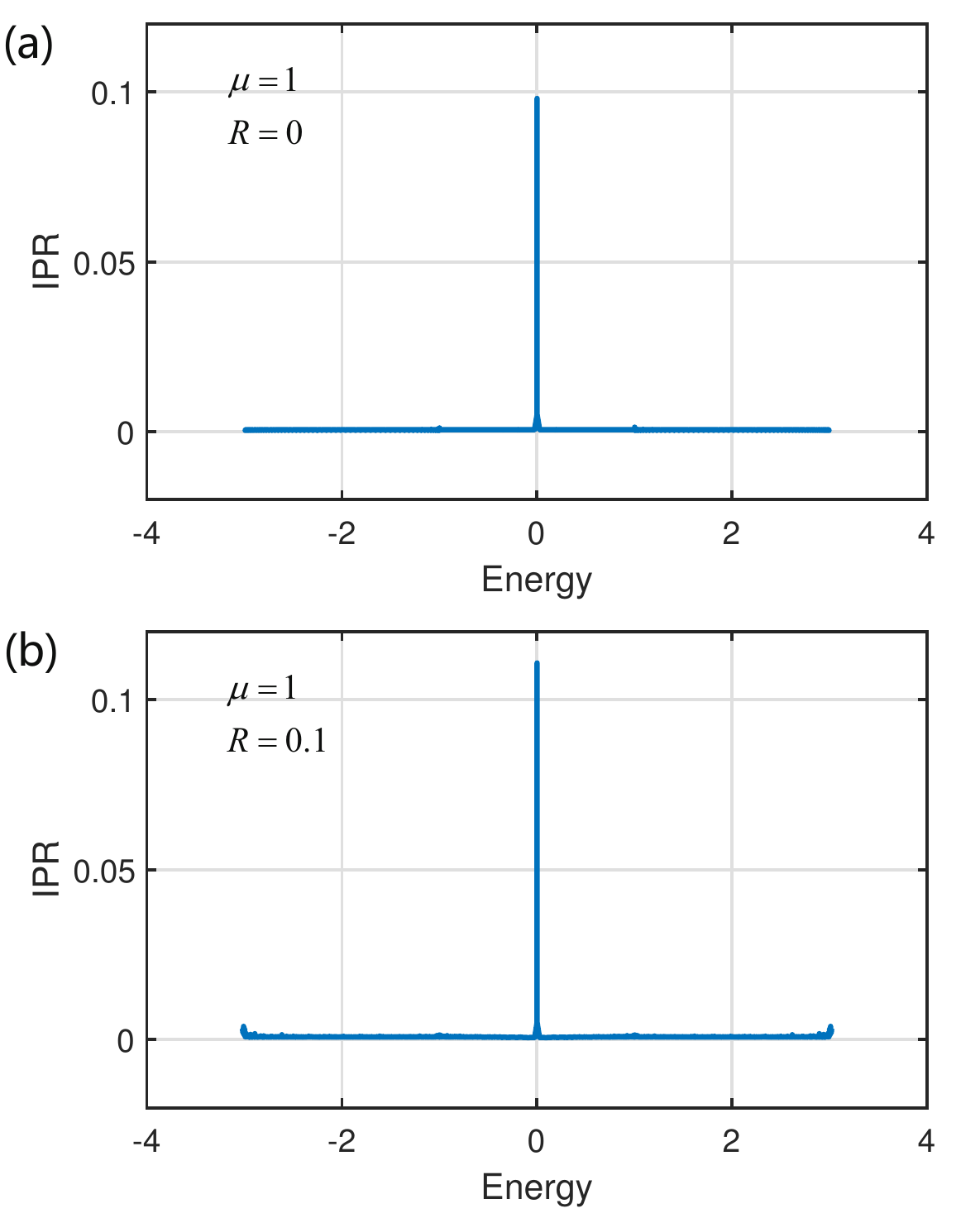}
\caption{Numerical results of inverse participation ratio (IPR)
for the gapless phase with $\protect\mu=1$ corresponding to Fig. \protect\ref%
{fig4}(a). (a) System without disorder. (b) System with
chiral-symmetry-preserving disorder $R=0.1$. The size of the system is $%
M\times N=60\times 60$.}
\label{fig5}
\end{figure}
which leads to $w=\pm 1$ for two degenerate points. The chiral relativistic
fermions serve as 2D Dirac points. Two Dirac points located at two separated
degenerate points have opposite chirality. We note that $w=0$ for $\mu =0$
and $\left\vert \mu \right\vert =2$. When $\mu =-2$ or $\mu =2$, two Dirac
points merge at $(0,0)$\ or $(\pm \pi ,\mp \pi )$ and become a single
degenerate point. The topology of the degenerate point becomes trivial, and
a perturbation hence can open up the bulk energy gap. We
illustrate the Bloch vector fields in $k_{x}$-$k_{y}$\ plane for several
typical cases in Fig. \ref{fig3}(a2)-(f2). As shown in figures, we find
three types of topological configurations: pair of vortices with opposite
chirality, single trivial vortex (or degeneracy lines), and no vortex,
corresponding to topological gapless, trivial gapless and gapped phases,
respectively. According to the bulk-boundary correspondence 
\cite{CKC, CKChiu2014}, the nontrivial bulk topology would leads to the
protected surface states and forming the flat band when the open boundary
condition is applied, as we can see in the following.

\subsection{Flat band edge modes}

Now we turn to study the feature of gapless phase of the square lattice. At
first, we revisit the description of the present model with cylindrical
boundary condition as shown in Fig. \ref{fig2}(a). Consider the Fourier
transformations in $y$ direction%
\begin{equation}
\left( a_{j,k_{y}},b_{j,k_{y}}\right) =\frac{1}{\sqrt{N}}%
\sum_{l=1}^{N}e^{-ik_{y}l}\left( a_{j,l},b_{j,l}\right) ,
\end{equation}%
where the wave vector $k_{y}=2\pi n/N$, $n=1,2,...,N$.\ The Hamiltonian $H$
can be rewritten as 
\begin{equation}
H=\sum_{k_{y}}H_{k_{y}},  \label{spectrum_H}
\end{equation}%
with 
\begin{equation}
H_{k_{y}}=\sum_{j=1}^{N}\delta _{k_{y}}a_{j,k_{y}}^{\dagger
}b_{j,k_{y}}+\sum_{j=1}^{N-1}a_{j,k_{y}}^{\dagger }b_{j+1,k_{y}}+\mathrm{%
h.c.,}  \label{SSH}
\end{equation}%
where $\delta _{k_{y}}=\left( \mu +e^{ik_{y}}\right) $, and $H_{k_{y}}$
obeys $\left[ H_{k_{y}},H_{k_{y}^{\prime }}\right] =0,$ i.e., $H$ has been
block diagonalized. We note that each $H_{k_{y}}$ %
represents a modified Su-Schrieffer-Heeger (SSH) chain with
hopping terms $\delta _{k_{y}}$ and $1$. The
schematic diagram is shown in Fig. \ref{fig2}(b).

The flat band edge modes of 2D chiral symmetric Hamiltonian Eq. (\ref{H})
with cylindrical boundary condition are originated from the zero energy edge
states of the modified SSH in Eq. (\ref{SSH}), which can be related to the
winding number \cite{RS,SMatsuura2013,Wong2013,MMili2017,JKABook} or
Zak phase \cite{PDelplace2011}. The winding number for the bulk Hamiltonian
of Eq. (\ref{SSH}) is defined as \cite{JKABook} 
\begin{equation}
\mathcal{W}\left( k_{y}\right) =\frac{1}{2\pi i}\int_{-\pi }^{\pi
}dk_{x}\partial _{k_{x}}\ln g\left( \mathbf{k}\right) ,
\end{equation}%
where $g\left( \mathbf{k}\right) $ is an off-diagonal element of the core
matrix $h(\mathbf{k})$ of the 2D bulk Hamiltonian in Eq. (\ref{2D_bulk}).
Direct derivation gives 
\begin{equation}
\mathcal{W}\left( k_{y}\right) =\left\{ 
\begin{array}{cc}
1, & \mu \left( \mu +2\cos k_{y}\right) <0 \\ 
0, & \mu \left( \mu +2\cos k_{y}\right) >0%
\end{array}%
\right. .  \label{winding}
\end{equation}
The winding number is $1$ for the parameters region $\mu \left( \mu +2\cos
k_{y}\right) <0,$ in which the open chain in Eq. (\ref{SSH}) is expected to
exist $1$ pairs of zero energy edge states \cite{MMili2017}, localized at
two ends of the chain, respectively. These zero energy edge states for all $%
k_{y}$ in the above parameters region form the flat band edge modes for the
2D lattice with cylindrical geometry.

One can always get a diagonalized $H_{k_{y}}$\ through the diagonalization
of the matrix of the corresponding single-particle SSH chain. Actually, it
can be checked that $H_{k_{y}}$ exits two zero modes in large $N$ limit 
\begin{equation}
\left\{ 
\begin{array}{c}
\left\vert \psi _{\mathrm{R}}\right\rangle =\Omega \sum_{j=1}^{N}\left(
-\delta _{k_{y}}^{\ast }\right) ^{N-j}a_{j,k_{y}}^{\dagger }\left\vert 
\mathrm{vac}\right\rangle \\ 
\left\vert \psi _{\mathrm{L}}\right\rangle =\Omega \sum_{j=1}^{N}\left(
-\delta _{k_{y}}\right) ^{j-1}b_{j,k_{y}}^{\dagger }\left\vert \mathrm{vac}%
\right\rangle%
\end{array}%
\right. ,
\end{equation}%
where $\Omega =\sqrt{1-\left\vert \delta _{k_{y}}\right\vert ^{2}}$ is
normalization constant, and $\left\vert \delta _{k_{y}}\right\vert <1$,
representing edge modes localizing at the right or left of the SSH chain.
The condition $\left\vert \delta _{k_{y}}\right\vert =\left\vert \mu
+e^{ik_{y}}\right\vert <1$ leads to $\mu \left( \mu +2\cos k_{y}\right) <0$,
and the interval of edge modes for $k_{y}$ is 
\begin{equation}
k_{y}\in \mathcal{I}=\left\{ 
\begin{array}{cc}
\left( -\pi ,-k_{y}^{\mathrm{c}}\right) \cup \left( k_{y}^{\mathrm{c}},\pi %
\right] , & 0<\mu <2 \\ 
\left( -k_{y}^{\mathrm{c}},k_{y}^{\mathrm{c}}\right) , & -2<\mu <0%
\end{array}%
\right. .
\end{equation}%
with $k_{y}^{\mathrm{c}}=\arccos \left( -\mu /2\right) .$ The above interval 
$\mathcal{I}$ matches with the interval with nonzero winding number in Eq. (%
\ref{winding}). The zero modes in the plot of energy band in Fig. \ref{fig3}%
(c3) and Fig. \ref{fig3}(e3) correspond this flat band of edge modes. For an
arbitrary site-state $a_{N,j}^{\dagger }\left\vert \mathrm{vac}\right\rangle 
$ (or $b_{1,j}^{\dagger }\left\vert \mathrm{vac}\right\rangle $) at the
edge, the total probability of on the component of edge state $\left\vert
\psi _{\mathrm{R}}\right\rangle $ (or $\left\vert \psi _{\mathrm{L}%
}\right\rangle $) is%
\begin{equation}
p=\frac{1}{N}\sum_{k_{y}\in \mathcal{I}}\left( 1-\left\vert \delta
_{k_{y}}\right\vert ^{2}\right) \approx \frac{1}{2\pi }\int_{\mathcal{I}%
}\left( 1-\left\vert \delta _{k_{y}}\right\vert ^{2}\right) dk_{y},
\end{equation}%
which is only $\mu $\ dependent in large $N$ limit. We will see that $p$ can
be measured by LE of the edge site-state.

\section{Dynamic detection of edge modes}

\label{Dynamic detection of edge modes}

\begin{figure}[tbp]
\centering
\includegraphics[width=0.45\textwidth]{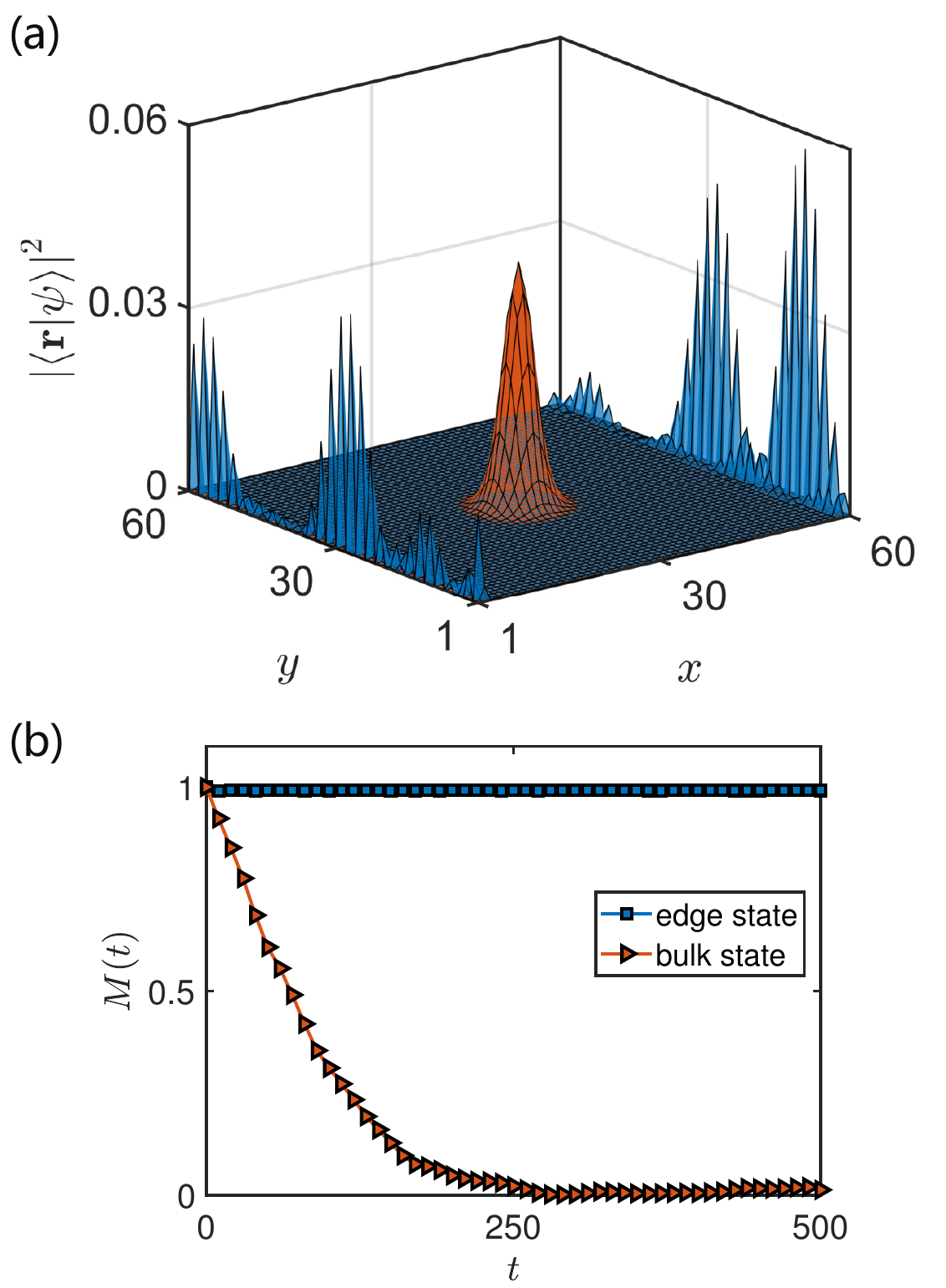}
\caption{(a) Profiles of initial states of numerical simulations for the
LEs. The edge state (blue) is taken as the eigenstate of the system in
cylindrical boundary condition without disorder and the bulk state (orange)
is taken as the 2D Gaussian wave-packet. (b) Plots of numerical simulations
for LEs as the functions of time. The initial states are taken as the edge
state and bulk state shown in (a). It can be seen that LEs have
diametrically opposite behaviors for the initial bulk and edge states. The
time $t$ is in units of $J^{-1}$, where $J$ is the scale of the Hamiltonian
and we take $J=1$. The size of the system is $M\times N=60\times 60$ and the
disorder strength is $R=0.1$.}
\label{fig6}
\end{figure}

In this section, we focus on the dynamics of the system in the presence of
disorder. As we know, one of the most striking features of topologically
protected edge states is the robustness against to certain types of
disordered perturbation to the original Hamiltonian. The disorder we discuss
here arises from the hopping integrals in the Hamiltonian $H$ from Eq. (\ref%
{H}) with cylindrical boundary condition. In the presence of disorder, the
Hamiltonian reads%
\begin{equation}
H_{\mathrm{D}}=\sum_{\mathbf{r}}\left( \mu _{\mathbf{r}}a_{\mathbf{r}%
}^{\dagger }b_{\mathbf{r}}+\nu _{\mathbf{r}}a_{\mathbf{r}}^{\dagger }b_{%
\mathbf{r}+\hat{x}}+\lambda _{\mathbf{r}}a_{\mathbf{r}}^{\dagger }b_{\mathbf{%
r}+\hat{y}}\right) +\text{h.c.,}
\end{equation}%
where parameters $\left\{ \mu _{\mathbf{r}},\nu _{\mathbf{r}},\lambda _{%
\mathbf{r}}\right\} $ are three set of position-dependent numbers. Here we
take 
\begin{equation}
\left\{ 
\begin{array}{c}
\mu _{\mathbf{r}}=\mu +d_{\mu ,\mathbf{r}} \\ 
\nu _{\mathbf{r}}=1+d_{\nu ,\mathbf{r}} \\ 
\lambda _{\mathbf{r}}=1+d_{\lambda ,\mathbf{r}}%
\end{array}%
\right. ,
\end{equation}%
where $d_{\mu ,\mathbf{r}},$ $d_{\nu ,\mathbf{r}},$ and $d_{\lambda ,\mathbf{%
r}}$ are uniform random real numbers within the interval $\left[ -R,R\right] 
$, taking the role of the disorder strength, and $\mathbf{r}$ is the site
index.

Now we investigate the influence of nonzero $R$%
\ by comparing two sets of eigenvalues obtained by numerical diagonalization
of finite-dimensional matrices of $H$ and $H_{\mathrm{D}}$ in
single-particle subspace, respectively. The plots in Fig. \ref{fig4}
indicate that the zero modes remain unchanged in the presence of
chiral-symmetry-preserving random perturbations with not too large $R$. The
chiral symmetry here is responsible for the existent of zero modes, in other
words, under chiral-symmetry-breaking disordered perturbation, the zero
modes no longer survive. Taking the disordered on-site potential for
example, the Hamiltonian reads $H^{\prime }=H+\sum_{\mathbf{r}}\left( d_{a,%
\mathbf{r}}a_{\mathbf{r}}^{\dagger }a_{\mathbf{r}}+d_{b,\mathbf{r}}b_{%
\mathbf{r}}^{\dagger }b_{\mathbf{r}}\right) $, where $d_{a,\mathbf{r}}$ and $%
d_{b,\mathbf{r}}$\ are uniform random real numbers within the interval $%
\left[ -R,R\right] $. The numerical results in Fig. \ref{fig4} indicate that
under this kind of chiral-symmetry-breaking disordered perturbation, the
zero modes do not survive, which may leads to the decay of the
LE in contrast to Eqs. (\ref{M0}) and (\ref{M1}) though the original edge
states remain localize in the edge. Furthermore, we investigate the inverse
participation ratio (IPR) for the gapless phase with and without
chiral-symmetry-preserving disorder. The IPR is defined as $\mathrm{IPR}%
\left( \mathrm{E}\right) =\sum_{\mathbf{r}}\left\vert \left\langle \mathbf{r}%
\right. \left\vert \psi _{\mathrm{E}}\right\rangle \right\vert ^{4}$, with $%
\mathrm{E}$ denotes the energy levels and $\mathbf{r}$ denotes the lattice
sites. The Numerical results of IPR shown in Fig. \ref{fig5} indicate that
all the states with energy $\mathrm{E}\neq 0$ are extended in the present or
absent of weak disorder, and the system is gapless in the transport sense.

According to the analysis in section \ref{Anderson localization and
Loschmidt echo}, the LEs should have diametrically opposite behaviors for
the initial bulk and edge states, respectively. To verify this point, we
compute the LEs for two initial states: (i) a Gaussian wave packet in the
bulk $\left\vert \psi _{\mathrm{G}}\right\rangle $; (ii) an edge state $%
\left\vert \psi _{\mathrm{R}}\right\rangle $\ or $\left\vert \psi _{\mathrm{L%
}}\right\rangle $. In Fig. \ref{fig6}, we plot the result, which is in
agreement with our prediction. We find that when $|\psi (0)\rangle $\ is a
bulk state $M(t)$\ will decay exponentially, while $M(t)$\ keeps in the
constant $1$\ when $|\psi (0)\rangle =\left\vert \psi _{\mathrm{R}%
}\right\rangle $\ or $\left\vert \psi _{\mathrm{L}}\right\rangle $. 
\begin{figure}[h]
\centering
\includegraphics[width=0.45\textwidth]{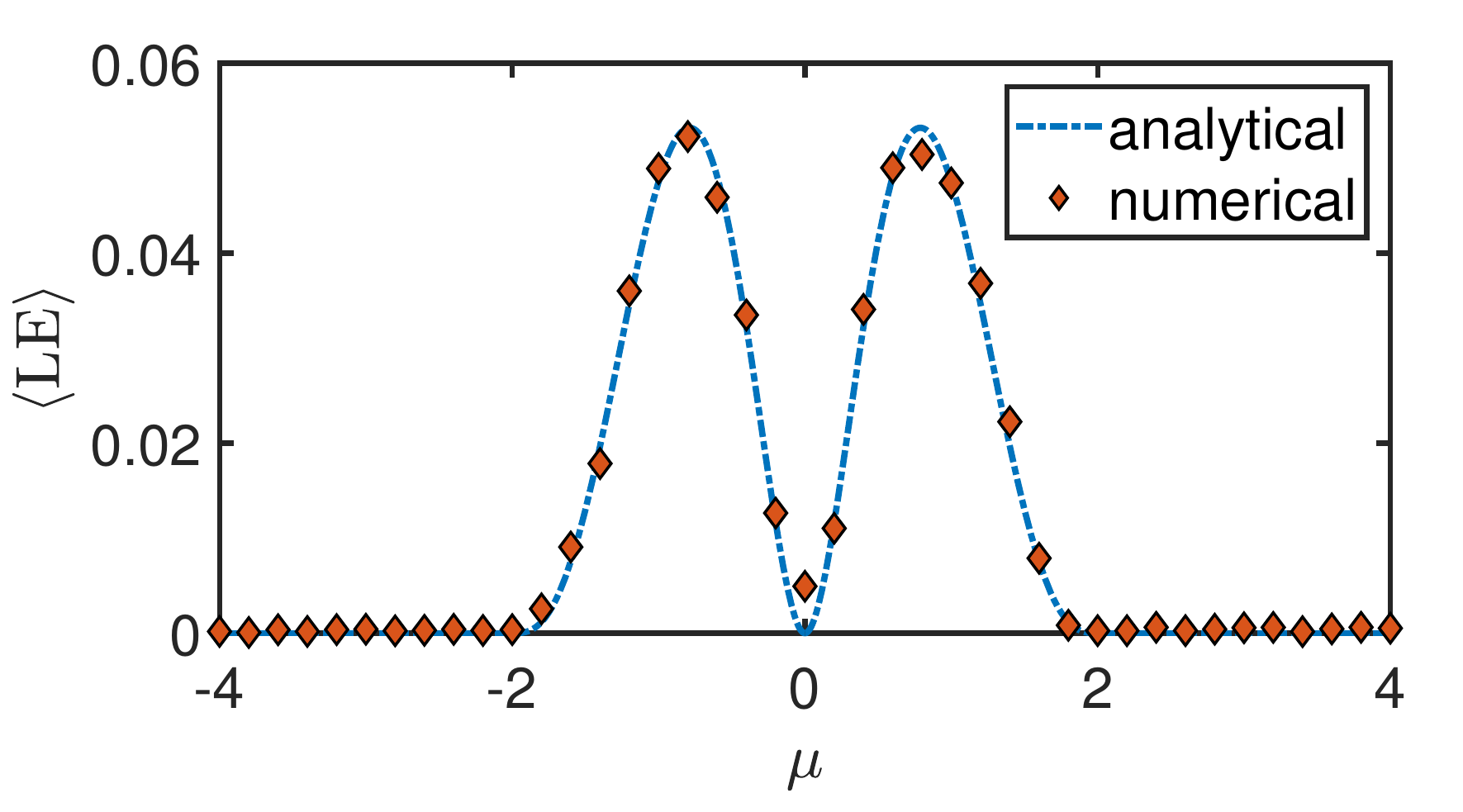}
\caption{Comparison of the average convergent LEs and analytical expression
from Eq.(\protect\ref{LE_analytical}) as the functions of $\protect\mu $.
The initial state is a edge site-state and the final time is $t=1000J^{-1}$,
where $J$ is the scale of the Hamiltonian and we take $J=1$. The size of the
system is $M\times N=60\times 60$ and the disorder strength is $R=0.1$. It
is found that the two results are in agreement with each other well. This
means that the measurement of LE can identify the phase diagram.}
\label{fig7}
\end{figure}
Accordingly, when we take the initial state as the superposition of
scattering and bound states, i.e.,%
\begin{equation}
|\psi (0)\rangle =c_{\mathrm{G}}\left\vert \psi _{\mathrm{G}}\right\rangle
+c_{\mathrm{R}}\left\vert \psi _{\mathrm{R}}\right\rangle +c_{\mathrm{L}%
}\left\vert \psi _{\mathrm{L}}\right\rangle ,
\end{equation}%
\ with $\left\vert c_{\mathrm{G}}\right\vert ^{2}+\left\vert c_{\mathrm{R}%
}\right\vert ^{2}+\left\vert c_{\mathrm{L}}\right\vert ^{2}=1$, we can have
the LE after long time%
\begin{equation}
\lim_{t\rightarrow \infty }M(t)=\left\vert c_{\mathrm{R}}\right\vert
^{2}+\left\vert c_{\mathrm{L}}\right\vert ^{2}=1-\left\vert c_{\mathrm{G}%
}\right\vert ^{2}.
\end{equation}%
It indicates that the magnitude of $c_{\mathrm{G}}$\ can be measured by the
LE.\ Furthermore, if we take $|\psi (0)\rangle =a_{N,j}^{\dagger }\left\vert 
\mathrm{vac}\right\rangle $\ (or $b_{1,j}^{\dagger }\left\vert \mathrm{vac}%
\right\rangle $), the population of survival zero modes is a function of $%
\mu $, which also relates to the quantity $p$, i.e.,%
\begin{equation}
\lim_{t\rightarrow \infty }M(t)=\lim_{t\rightarrow \infty }\left\vert
\langle \psi (0)|e^{iH_{\mathrm{D}}t}e^{-iH_{0}t}|\psi (0)\rangle
\right\vert ^{2}\approx p^{2},  \label{LE_analytical}
\end{equation}%
for very weak disordered system $H_{\mathrm{D}}$. It is presumably that the
size of flat band $k_{c}$\ can be obtained by the LE in the dynamical
process.

To demonstrate and verify this scheme, we perform numerical simulations. We choose three different strengths of
chiral-symmetry-preserving disorder $R$ and six typical values of hopping
amplitudes $\mu$. The numerical simulations are performed ten times for each
set of parameter. Fig. \ref{fig3}(a4-f4) plot the convergent LEs, where 
\textrm{LE}$=\lim_{t\rightarrow \infty }M(t)$ is obtain by
taking a sufficiently large $t$ ($t=3000$), for several typical $\mu$ with
different strengths of chiral-symmetry-preserving disorder $R=0.04$, $0.1$
and $0.15$. It indicates that a single measurement result depends on the
setting random number. The average of multi-measurement result $\left\langle 
\mathrm{LE}\right\rangle $ is very close to the analytical result in the blue dashed lines. The dependence of $\left\langle \mathrm{%
LE}\right\rangle $ on $\mu $ for wide range of $\mu $ with the
disorder strength $R=0.1$ are presented in Fig. \ref{fig7}. The comparison
between analytical and numerical results show that the LE method has a good
accuracy to determine the positions of vortices, as well as the phase
diagram. The transition points occurs at $\mu =\pm 2$, associated with the
vanishing $\left\langle \mathrm{LE}\right\rangle $.

The data and codes of the numerical calculations of Figs. 3-7 are available
in supplementary material as well as in Zenodo \cite{data}.

\section{Discussion}

\label{Discussion} In this work, we have proposed a way to detect the
positions of two vortices in 2D momentum space, as well as the phase
diagram. The advantage of this scheme is not limited by the imperfection of the system, but in the aid of the disorder. Photonic system is an candidate for the realization of the scheme in experiment, beyond the solid-state electron systems. The field of topological photonics grows
rapidly and aims to explore the physics of topological phases of matter in
the context of optics. Photonic systems provide a natural and convenient
medium to investigate fundamental quantum transport properties. Using
photons, one can selectively excite a site-state, and observe the spatial
responses throughout the material, which are challenging tasks in electronic
systems.\ Recently, it has be shown that Loschmidt echo of photons can be
observed in a binary waveguide, by exchanging the two sublattices after some
propagation distance \cite{SLonghi}. The dynamic feature of topological edge
states and phase diagram presented in this work potentially can be utilized
for developing inherently robust artificial photonic devices.

\section*{Acknowledgment}

This work was supported by National Natural Science Foundation of China
(under Grant No. 11874225).

\end{document}